\documentclass[iop,natbib]{emulateapj}
\usepackage{hyperref}
\hypersetup{
  colorlinks   = true,    
  urlcolor     = blue,    
  linkcolor    = blue,    
  citecolor    = blue      
}

\usepackage{natbib}
\usepackage{xcolor}

\shorttitle{Holey Debris Disks: APP}
\shortauthors{Meshkat et al.}

\slugcomment{Accepted for publication in ApJ} 

\begin{document}

\title{Searching for Planets in Holey Debris Disks with the Apodizing Phase Plate}

\author{Tiffany Meshkat$^{1,2}$, Vanessa P. Bailey$^3$,  Kate Y. L. Su$^3$, Matthew A. Kenworthy$^2$,  Eric E. Mamajek$^4$, Philip M. Hinz$^3$, Paul S. Smith$^3$}
\altaffiltext{1}{Based on observations collected at the European
  Organization for Astronomical Research in the Southern Hemisphere,
  Chile, ESO under program numbers 090.C-0148(A) and 091.C-0457(A)}
\altaffiltext{2}{Sterrewacht Leiden, P.O. Box 9513, Niels Bohrweg 2,
  2300 RA Leiden, The Netherlands}
\altaffiltext{3}{Steward Observatory, Department of Astronomy, University 
of Arizona, 933 North Cherry Avenue, Tucson, AZ 85721-0065, USA}
\altaffiltext{4}{Department of Physics and Astronomy, University of 
Rochester, Rochester, NY 14627-0171, USA}


\begin{abstract}
We present our first results from a high-contrast imaging search for 
planetary mass companions around stars with gapped debris disks, as inferred from 
the stars' bright infrared excesses. For the six considered stars, we model the disks' 
unresolved infrared spectral energy distributions (SEDs) in order to derive 
the temperature and location of the disk components. With VLT/NaCo Apodizing Phase Plate 
coronagraphic $L'$-band imaging, we search for planetary mass companions that may be sculpting 
the disks. We detect neither disks nor companions in this sample, confirmed
by comparing plausible point sources with archival data.
In order to calculate our mass sensitivity limit, we revisit the stellar age estimates. 
One target, HD 17848, at 540$\pm$100 Myr old is significantly older than previously estimated. 
We then discuss our high-contrast imaging results with respect to the disk properties. 

\end{abstract}

\section{Introduction}

Despite the large number of direct imaging surveys (SEEDS: \citealt{Brandt14,Janson13}, NICI:  
\citealt{Biller13,Nielsen13,Wahhaj13}, GDPS: \citealt{Lafreniere07}, IDPS: \citealt{Vigan12}, 
and with NaCo: \citealt{Desidera14,Chauvin14})
and the hundreds of sources observed, few planet have been discovered from these surveys
(HR8799 bcde from IDPS: \citealt{Marois08, Marois10}, GJ 504 b from SEEDS; \citealt{Kuzuhara13}). 
This low detection rate is likely due to a combination of factors 
(e.g. the wavelength of the observations, the target selection, the lack of a dedicated 
exoplanet instrument, etc). The faint $H$-band detection of HD 95086 b \citep{Galicher14} 
and the non-detection companion candidate around HD 169142 \citep{Reggiani14,Biller14} demonstrate 
that these low mass companions are extremely red. These results reinforce the importance of searching for planets in 
the thermal infrared with $L'$-band (3.8 $\mu$m), as the planet-to-star contrast is reduced and 
contamination from background stars is strongly suppressed.

We aim to find a signpost for planets; a trait of the stars or systems 
which might yield a higher planet detection rate than previous surveys.
One possible signpost is the structure of debris disks with gaps, which
dynamically may imply the presence of a planet \citep{Quillen02, Quillen06}. 
This includes both debris systems with a large hole in the center,  
generally classified as one-component disks, and debris systems with 
a large gap, generally classified as two-component (warm inner and cool outer) disks \citep{Su14}.
We have designed the ``Holey Debris Disks'' exoplanet direct imaging survey guided by this hypothesis. 
The targets were selected based on several criteria: 
youth, distance, brightness, and unresolved infrared excess indicative of the 
presence of a possibly-sculpted debris disk around the star using Spitzer 
data (Su et al. 2010). Our constraints on companion masses and locations will provide useful 
inputs for future debris disk/planet dynamical models.

We present the first results of the Holey Debris Disk survey, obtained at the Very Large Telescope (VLT) 
with the NaCo\footnote{Nasmyth Adaptive Optics System (NAOS) Near-Infrared Imager and Spectrograph (CONICA)} 
(\citealt{Lenzen03,Rousset03}) instrument and Apodizing Phase Plate coronagraph (APP: \citealt{Kenworthy10b, Quanz10}). 
The remainder of the survey data, obtained with the LMIRCam\footnote{L/M-band mid-InfraRed Camera} \citep{Skrutskie10,Leisenring12}, 
Clio2 \citep{Sivanandam06}, and NICI\footnote{Near-Infrared Coronagraphic Imager} \citep{Chun08} instruments, will be presented in a companion paper.
Subsets of the SEEDS survey \citep{Janson13} and the NICI survey 
\citep{Wahhaj13} focused on similar debris disk 
targets at $H$-band; however, we chose to image our sample with NaCo in the thermal infrared 
($L'$-band). This method does not rely on methane absorption in the planet's 
atmosphere (as Spectral Differential Imaging does; \citealt{Marois00}), nor is it negatively 
impacted by the reddening effect of thick clouds. The majority of planets 
found to date, despite their relatively cool effective temperatures, lack 
methane absorption and have thick clouds (eg: HR8799 bcd and 2MASS 1207 b \citealt{Skemer14}).

The direct detection of close-in planets is limited by 
instrumental diffraction and scattering effects on the point spread function (PSF) of the 
bright primary star. The scattered light may be much 
brighter than a companion. Pupil apodizing coronagraphs block the primary star's light, 
suppressing its PSF at the expense of throughput ($\sim 40\%$ suppressed for the APP: \citealt{Quanz10}). 
We use the APP coronagraph on NaCo at the VLT to increase our sensitivity 
around our stars. 

This is the first paper in a series for the Holey Debris Disks project, discussing the VLT/NaCo APP coronagraphic 
data and results. In Section \ref{sec:obs} we describe the APP observations and data reduction.
In Section \ref{sec:disk_properties} we present the methods and data used 
for deriving the disk properties. 
In Section \ref{sec:discussion} we show the resulting contrast curves for each 
of our targets and discuss the significance of our sensitivity with respect 
to the disk properties derived from disk models. We conclude in Section \ref{sec:conclusions}.

\section{APP Observations and Data Reduction}
\label{sec:obs}
\subsection{Observations}
\begin{table*}[htb]
\scriptsize
\caption{Observing Log for APP NaCo/VLT runs 090.C-0148(A) and 091.C-0457(A).} 
\centering 
\begin{tabular}{l l l l l}
\hline
   Target & Observation dates\footnote{Data are listed in chronological order from when the first hemisphere was observed. The last two
targets were only observed in one APP hemisphere. Any difference in integration time 
between hemispheres was unintentional, simply due to the conditions of the night.} UT (Hem 1, Hem 2) & Total integration time (s) & On-sky rotation ($^{\circ}$)  & Average DIMM seeing ($\arcsec$) \\ [0.5ex]
  \hline\hline
  
   HD 28355 & 2012 Nov 17, 2013 Jan 14 & 2580, 2060  & 24.80, 19.56 & 0.57, 0.78  \\ 
      
   HD 17848 & 2012 Nov 22, 2013 Jan 15 & 1940, 2860 & 19.37, 28.55  & 0.755, 0.89 \\  
      
   HD 37484 & 2013 Jan 21, 2013 Jan 25 & 1360, 2232 & 27.86, 10.82 & 1.16, 0.6 \\
   
   HD 95086 & 2013 Apr 26, 2013 May 16 & 3800, 3120 & 24.85, 20.50 & 0.76, 1.4 \\
         
   HD 134888 & 2013 Apr 21, -- & 3800, -- & 93.04, -- & 1.22, -- \\

   HD 110058 & 2013 Apr 25, -- & 3250, -- & 33.53, -- & 1.425, -- \\
 
\hline

   \end{tabular}  
\label{table:data}

\end{table*}

The APP data were obtained in 2012 and 2013  
(090.C-0148(A) and 091.C-0457(A), PI: Tiffany Meshkat) 
at the VLT/UT4 with NaCo.
The APP was used for additional diffraction suppression from $0\farcs2$ to
$1\farcs0$, increasing the chance of detecting faint companions close to the 
target stars. The infrared wavefront sensor was used with the target stars themselves as 
the natural guide star. Data were acquired with the L27 camera (27 mas/pix) and 
the $L'$-band filter ($\lambda$ = 3.80$\mu$m and $\Delta\lambda$\,=\,0.62$\mu$m).
We used pupil tracking mode for Angular Differential Imaging (ADI; \citealt{Marois06})
and intentionally saturated the stellar point-spread-function (PSF) core 
to increase the signal-to-noise (S/N) from possible faint companions.
We also obtained unsaturated data to calibrate the photometry relative 
to the central star and determine the sensitivity achieved in each dataset.

The APP suppresses diffraction over 
a 180$^{\circ}$ hemisphere on one side of the target star. 
Thus, two datasets need to be acquired, with different initial position angles
(P.A.) for full coverage around the target star.
We observed six targets with the APP (see \autoref{table:data}), however only four of 
these have complete APP hemisphere coverage around all P.A.s. One of the  
targets (HD 134888) has 270\arcdeg\ coverage with only one APP hemisphere (from 
135\arcdeg\ to -135\arcdeg), the final 
target (HD 110058) does not have sufficient coverage for a detailed analysis. 

Data were obtained in cube mode. Each data cube contains 200 frames, with 
an integration time between 0.1 s to 0.3 s per frame, depending on the 
$L'$ mag of the star. Details about total integration time 
and field rotation per target are contained in \autoref{table:data}. 
A three-point dither pattern (with an amplitude of 4$\arcsec$) was used on the detector to subtract the 
sky background systematics from each dataset, detailed in \citet{Kenworthy13}.

\subsection{Data Reduction}
\label{sec:data_reduction}
Data cubes closest in time at different dither positions were subtracted 
from each other and centroided. The average is taken over the subtracted 
cube in order to decrease the full dataset size. The final averaged frames 
cover a rectangular area of $3\farcs1$ by $1\farcs5$ centered on the star. 
The reason for this asymmetry is that the APP can only observe on the upper quartile portion of the CCD, 
due to the wedge, deliberately introduced in the optics to avoid ghost images \citep{Kenworthy10b}. 
Thus, we have complete coverage around the star out to $1\farcs5$ and incomplete coverage beyond 
that radius.

Optimized principal component analysis (PCA) was run on each target APP hemisphere independently, 
following \citet{Meshkat14}. PCA processed frames were derotated, averaged over, 
and combined with the other hemisphere to generate the final image with North facing up.
If two regions of APP data overlap, those regions were combined by average.
We generated final PCA processed frames for a range of principal components (PCs) from 3 up to 20. 
This first test was to determine if there were any companion candidates in our data, 
and if so, how robust they were to the number of PCs used in the image reduction. 

We next fixed the number of PCs (approximately 10\% of cubes in the dataset). 
We injected artificial planets in the data cubes
and ran the extraction pipeline to generate contrast curves of point sources sensitivity 
for 5$\sigma$. The artificial planets were generated 
from the data with unsaturated PSF cores. The unsaturated data was scaled 
to the same exposure time as the saturated data. Since the APP is in the 
pupil plane, it affects the PSF of every source in the FOV in the same way, thus 
we can use the unsaturated star itself to generate artificial planets.
The artificial planets were added to the data before PCA. The planets 
were added at different angular separations from $0\farcs18$ to $3\farcs0$ in steps of $0\farcs15$ 
and at different contrasts from 5 to 12 mag in steps of 1 mag. This was 
repeated for two different P.A.s, in order to place a fake planet in each 
APP hemisphere. We smoothed the final PCA image by a 1 $\lambda$/D aperture \citep{Bailey13}. 
We measured the S/N of the injected planet and decreased the flux of the injected 
planet until it reached a S/N of 5. The S/N was determined by 
dividing the flux in one pixel at the center of the injected planet by 
the noise in a 2 pixel wide ring around the star at the planet separation
(not including the planet). 
We then interpolated between the contrasts to determine the 5$\sigma$ contrast limit.

\section{Debris Disk SEDs and Derivation of Disk properties}
\label{sec:disk_properties}

For the five stars that have complete or nearly complete APP
coverage, we derive disk properties using both {\it Spitzer} and {\it Herschel}
data with SED models and quote the results of HD 95086 from Su et al. (in press). 
We focus on the properties of the cold component, because
it is the dominant part of the debris disk SED and is also
more relevant to our direct imaging observations for 
low-mass companions because of the inner working angle in our images.

\begin{figure*}
\plottwo{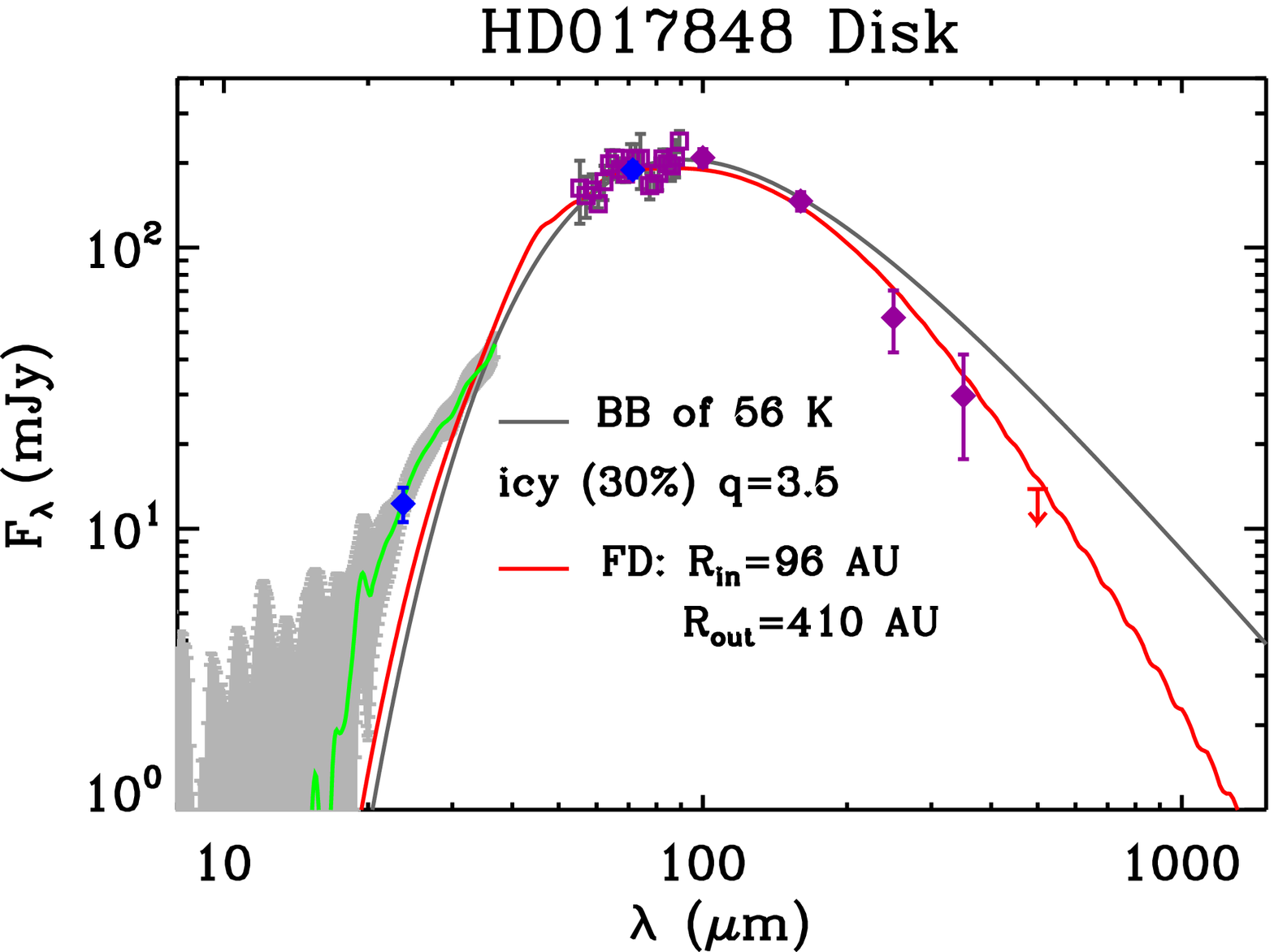}{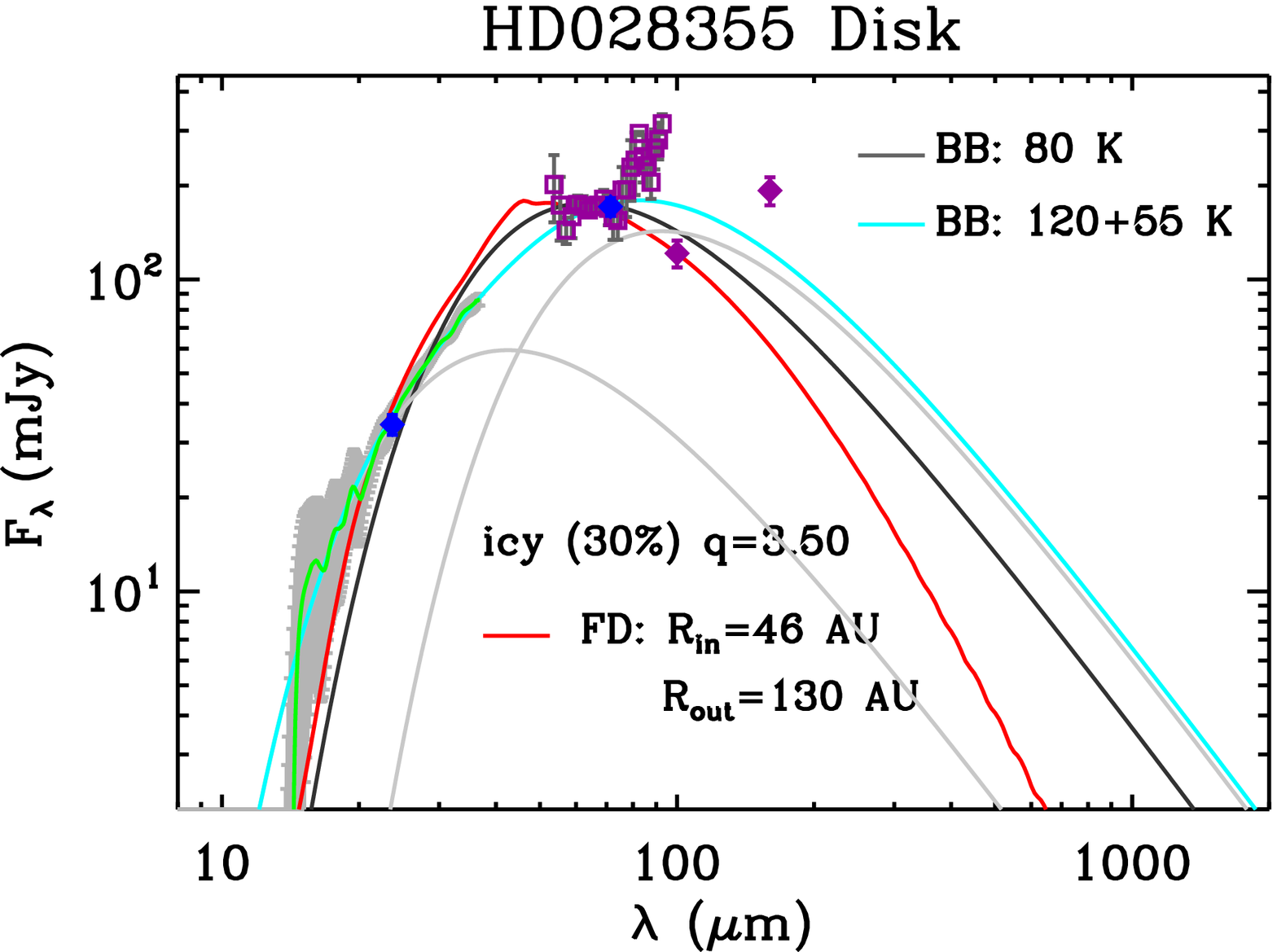}\\
\plottwo{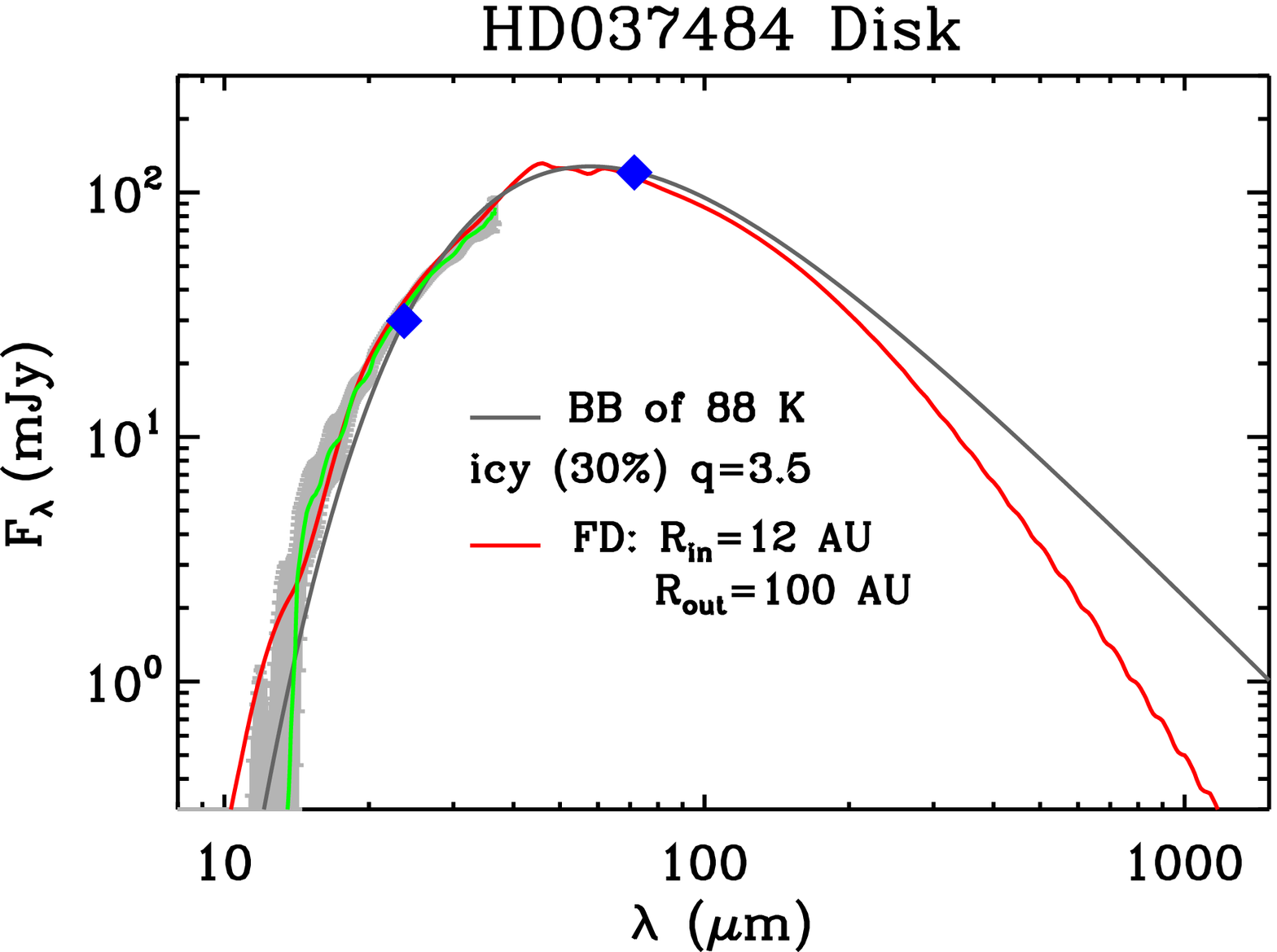}{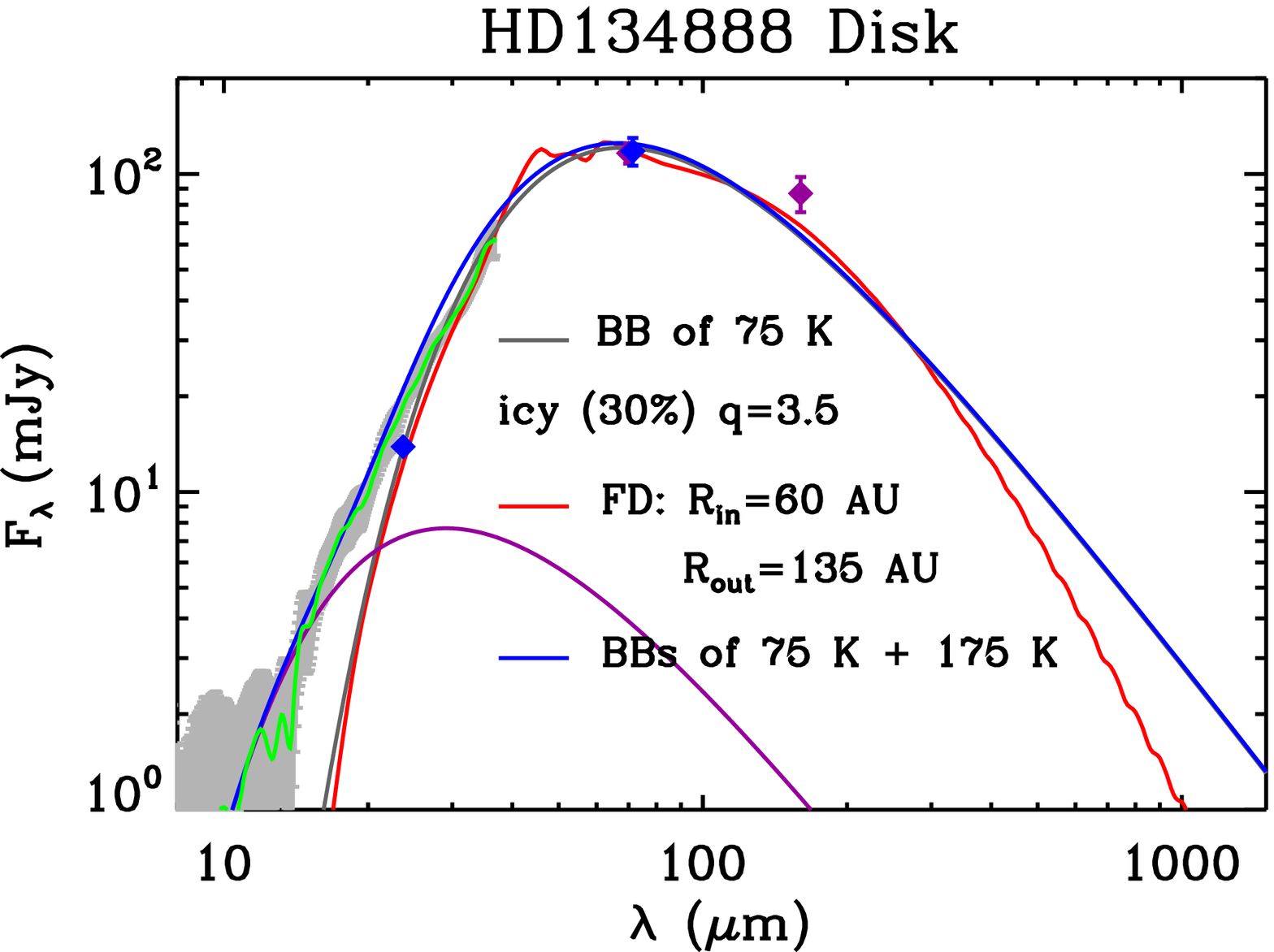}\\
\caption{The SEDs of the four debris systems used to derive
disk properties. For all the targets, the stellar photosphere has been subtracted.
In all panels, the observed data are shown in various
symbols and colors: blue diamonds: \textit{Spitzer} MIPS 24 and 70
$\mu$m photometry; purple diamonds: \textit{Herschel} PACS and SPIRE
photometry; small green dots: \textit{Spitzer} IRS spectrum; and
purple squares: \textit{Spitzer} MIPS-SED data. Various lines are SED fits:
black solid lines for single blackbody emission, and red solid lines for the 
one-component icy grain SED model. For both the HD 28355 and HD 134888 disks, two blackbody temperature fit is 
also shown for comparison.
}
\label{fig:SEDs}
\end{figure*}

\subsection{Spitzer and Herschel data reduction}
\label{subsec:ancillary_data}
Our targets were selected based on the presence of a bright, unresolved 
infrared excess indicative of a dusty debris disk. Most of them have well covered SEDs 
using \textit{Spitzer} broad-band and spectroscopic data and
\textit{Herschel} broad-band photometry. We collected all the
published broad-band photometry from the literature, and performed our own
photometry measurements if the data were not published. \textit{Spitzer}
MIPS 24 and 70 $\mu$m photometry is part of the \textit{Spitzer}
Debris Disk Master Catalog (Su et al.\ in prep.), where data reduction
and photometry extraction were briefly summarized in
\citet{Sierchio14}. \textit{Herschel} PACS data reduction and
photometry extraction were performed following the procedure published
by \citet{Balog14} for calibration stars, except that, as detailed in the 
following section, smaller apertures
were used for photometry measurements to minimize background
contamination. All our debris disk targets have existing
\textit{Spitzer} IRS low-resolution spectra. We downloaded the
extracted spectra using the CASSIS database
\citep{Lebouteiller11}. Three of the targets have MIPS-SED
low-resolution spectra covering 55 to 95 $\mu$m with a slit width of
$\sim$20\arcsec. We reduced and calibrated the MIPS-SED data as
described by \citet{Lu08}.

\subsection{Spitzer and Herschel fluxes}
\textit{Herschel} photometry of HD 95086 has been published by
\citet{Moor13} and Su et al.\ (in press). Here we briefly summarize
the {\it Hershel} photometry results for HD 134888, HD 28355 and
HD17848. The PACS 70 and 160 $\mu$m observation for HD 134888 was
obtained under Program {\it OT1\_dpadget\_1}. The source appears to be
point-like surrounded by background cirrus structure apparent on the
160 $\mu$m data, therefore we used an aperture size of 6\arcsec\ at 70
$\mu$m and 11\arcsec\ at 160 $\mu$m to measure the
photometry. Including the absolute calibration errors (7\%,
\citealt{Balog14}), the PACS photometry for HD 134888 is: 117$\pm$8.4
mJy and 87$\pm$11 mJy at 70 and 160 $\mu$m, respectively, and the PACS
70 $\mu$m photometry agrees with the MIPS 70 $\mu$m photometry
(119$\pm$12 mJy) very well.

For HD 28355, PACS 100 and 160 $\mu$m data were obtained under Program
{\it OT2\_fmorales\_3}. The source appears to be point-like, located near a
bright source $\sim$28\arcsec\ away. To avoid contamination from the
nearby bright source which is extended at 160 $\mu$m, we used an
aperture size of 6\arcsec\ to measure the photometry at both
bands. Our adopted photometry for the HD 28355 system is 127$\pm$12
mJy and 195 mJy$\pm$20 mJy at 100 and 160 $\mu$m, respectively. The
contamination from the nearby bright source explains the up-turn in
the MIPS-SED data for wavelengths longward of $\sim$70 $\mu$m (see
\autoref{fig:SEDs}). 

The PACS and SPIRE data for HD 17848 were obtained under program
{\it OT1\_pabraham\_2} using all the available photometry bands. The source
appears to be elongated, suggesting a close to edge-on orientation to
the disk. We used an aperture size of 22\arcsec\ to measure photometry
at all three bands, at which radius the encircled flux reached the
maximum and flattened afterwards. The final PACS fluxes are: 201$\pm$14
mJy, 213$\pm$15 mJy, and 148$\pm$11 mJy at 70, 100 and 160 $\mu$m,
respectively. The PACS 70/100 $\mu$m photometry agrees well with the
MIPS 70 $\mu$m and MIPS-SED data (see \autoref{fig:SEDs}). For the
SPIRE data, we used the level 2 product for point sources provided by
the \textit{Herschel} Science Center (HIPE ver.\ 11). The source was
detected at $\gtrsim$3 $\sigma $ at both 250 and 350 $\mu$m, but not
detected at 500 $\mu$m. The submillimeter fluxes are: 57$\pm$14 mJy,
30$\pm$12 mJy and $<$14 mJy (1$\sigma$) at 250, 350 and 500 $\mu$m,
respectively. These values are consistent with the {\it Herschel}
photometry recently published by \citet{Pawellek14} within
uncertainties.

The final disk SEDs were constructed by subtracting the best-fit
Kurucz ATLAS9 stellar atmosphere models \citep{Castelli04} that fit the
optical and near-infrared data. The uncertainties in the excess fluxes
also included 2\% photospheric extrapolation
errors. \autoref{fig:SEDs} shows the disk SEDs for our debris
targets. We have excluded the disk SED for HD 95086, which will be
published in Su et al.\ (in press), and for HD 110058, on which we did not
obtain sufficient APP sky coverage for analysis. 

\subsection{Methodology of Deriving Disk Properties}
\label{sec:methodology}

To estimate the likely debris location, we performed basic SED modeling. We started
with the simplest blackbody fitting for the disk SEDs (with a typical
error of $\pm$5 K) and used these temperatures to guide a more
complicated SED model with adopted grain properties. Without spatial
information, SED modeling is degenerate; therefore, our strategy is to
minimize model parameters with some reasonable assumptions. 

Similar to the SED model for HD 95086 (Su et al.\ in press), we
adopted icy silicates as our grain properties. The particle size distribution was assumed to be a power-law
form, $\sim a^{-q}$, where $a$ is the grain radius with a minimum
$a_{min}$ and maximum $a_{max}$ cutoffs. We adopted $q=3.5$ for the
power index of the particle distribution and $a_{max}=1000~\mu$m for
all the models. The minimum grain size is set to be close to the
radiation blowout size estimated based on the grain density and 
stellar luminosity and mass. We assumed that the debris
is uniformly distributed (constant surface density) from the inner
radius ($R_{in}$) to the outer radius ($R_{out}$), and computed the
thermal dust emission under optically thin conditions where the star
is the only heating source. We then derived the best-fit inner and outer 
boundaries of the cold disk component, along with the total 
cold disk dust mass, quantifying the goodness of fit with reduced $\chi^2$. As detailed in the following 
sections, we excluded any long wavelength data which was contaminated by nearby background 
galaxies, as well as any short wavelength data which might include a contribution from a warm 
disk component.

In some cases, weak warm excess shortward of $\sim20\mu m$ was present. When 
the warm excess signal was greater than the uncertainty in photospheric subtraction, 
we also performed a blackbody fit to the short wavelength data to derive the warm 
component's approximate temperature. Because the location of the tentative warm 
disks ($\lesssim$2 to 10 AU) is less than the inner working angle of our high contrast 
observations, we do not perform detailed SED modeling for the warm excesses. We 
comment on the derived warm dust temperatures, but their nature will be discussed 
in a separate paper.

\section{Results and Discussion}
\label{sec:discussion}

\begin{table*}[htb]
\scriptsize
\caption{Stellar and disk properties for targets.} 
\centering 
\begin{tabular}{l l l l l l l }
\hline
   Target  & Distance (pc) & $L'$ (mag) & Age (Myr) & Cold disk temperature (K) & Cold disk inner/outer edge (AU) & Dust mass ($10^{-3} M_{\earth}$) \\
  \hline\hline
   HD17848& 50.5$\pm$0.5 & 5.0 & 540 & 56 & 96$^{+9}_{-37}$, 410$^{+24}_{-96}$ & 1.3$\pm$0.7\\ 
   
   HD28355& 48.8$\pm$0.7 & 4.5 & 625 & 80 & 46$\pm$12, 130$\pm$30 & 1.8$\pm$0.7\\ 

   HD37484& 56.7$\pm$2.0 & 6.3 & 30 & 88 & $12^{+20}_{-4}$, $100^{+100}_{-20}$ & 2$\pm$1\\ 
   
   HD95086& 90.4$\pm$3.3 & 6.7 & 17 & 55 & 63$\pm$6, 189$\pm$13 & 200 \\ 
 
   HD134888& 89.0$\pm$8.4 & 7.6 & 16 & 75 & 60$\pm$11, 135$\pm$29 & 32.5$\pm$14\\   
   
   HD110058& 106.7$\pm$8.3 & 7.5 & 10 & --- & --- & ---\\      
   \hline   
   \end{tabular}  
\label{table:properties}

\end{table*}

\begin{figure*}
\plottwo{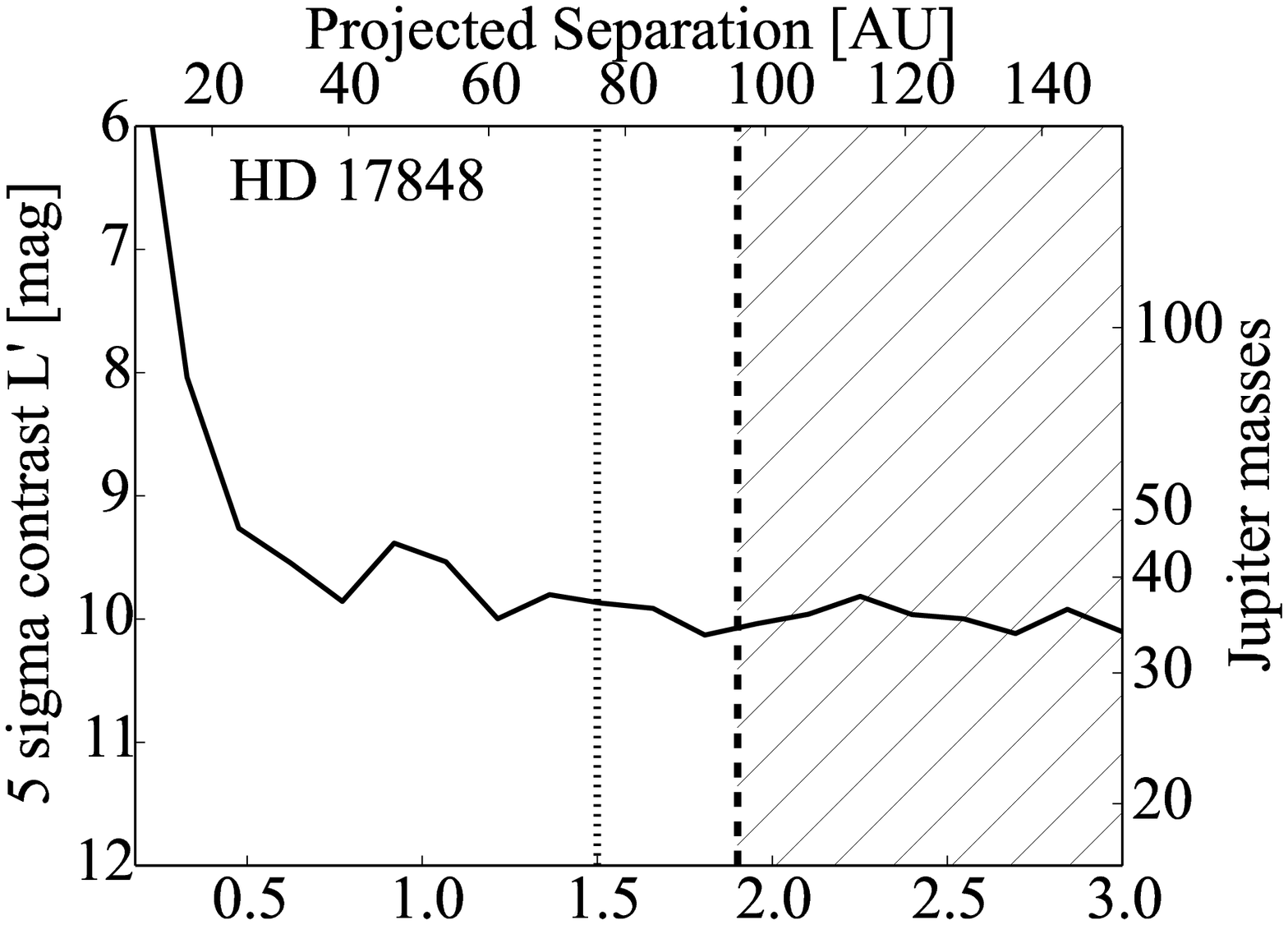}{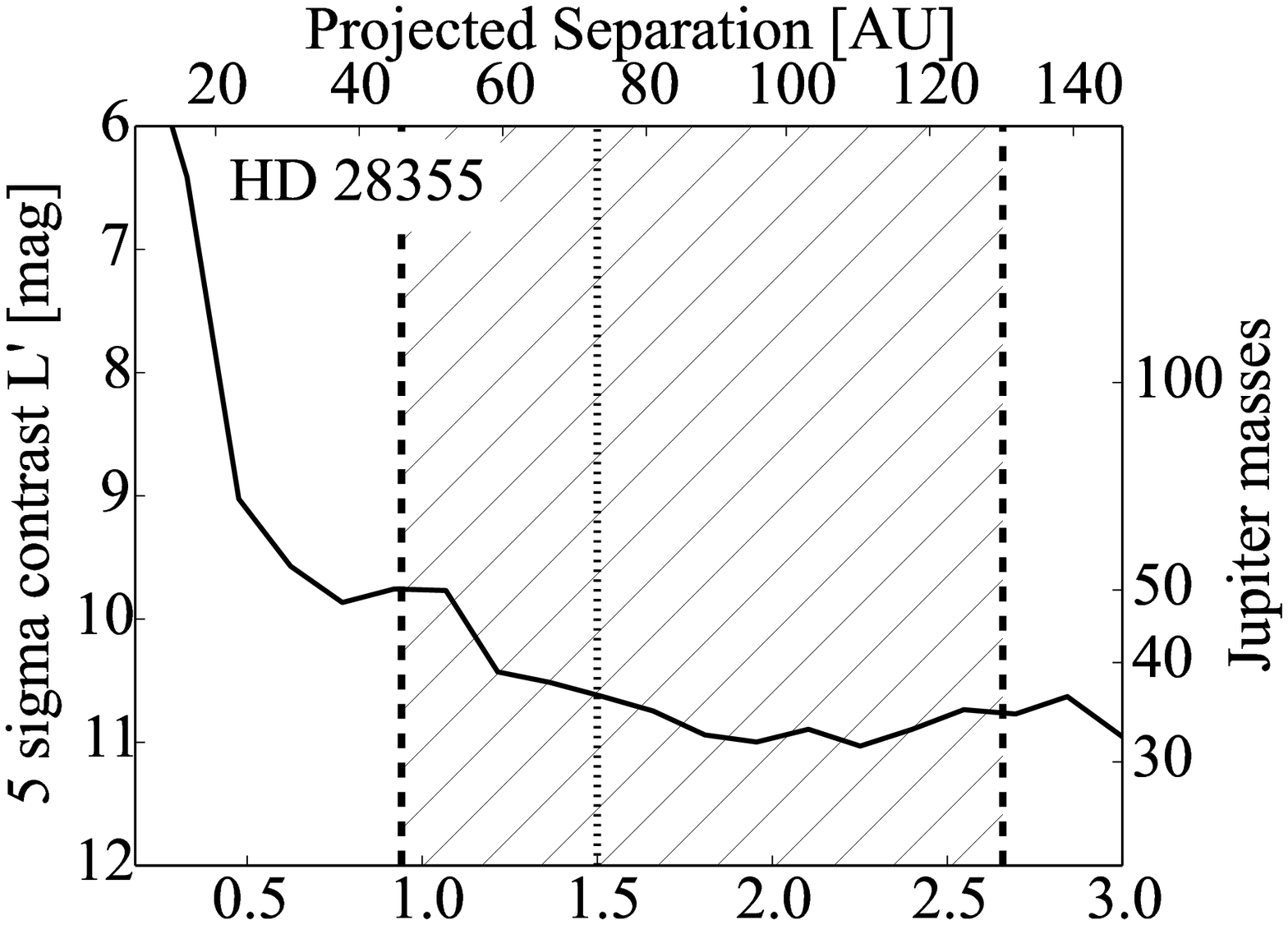}\\
\vspace{10pt}
\plottwo{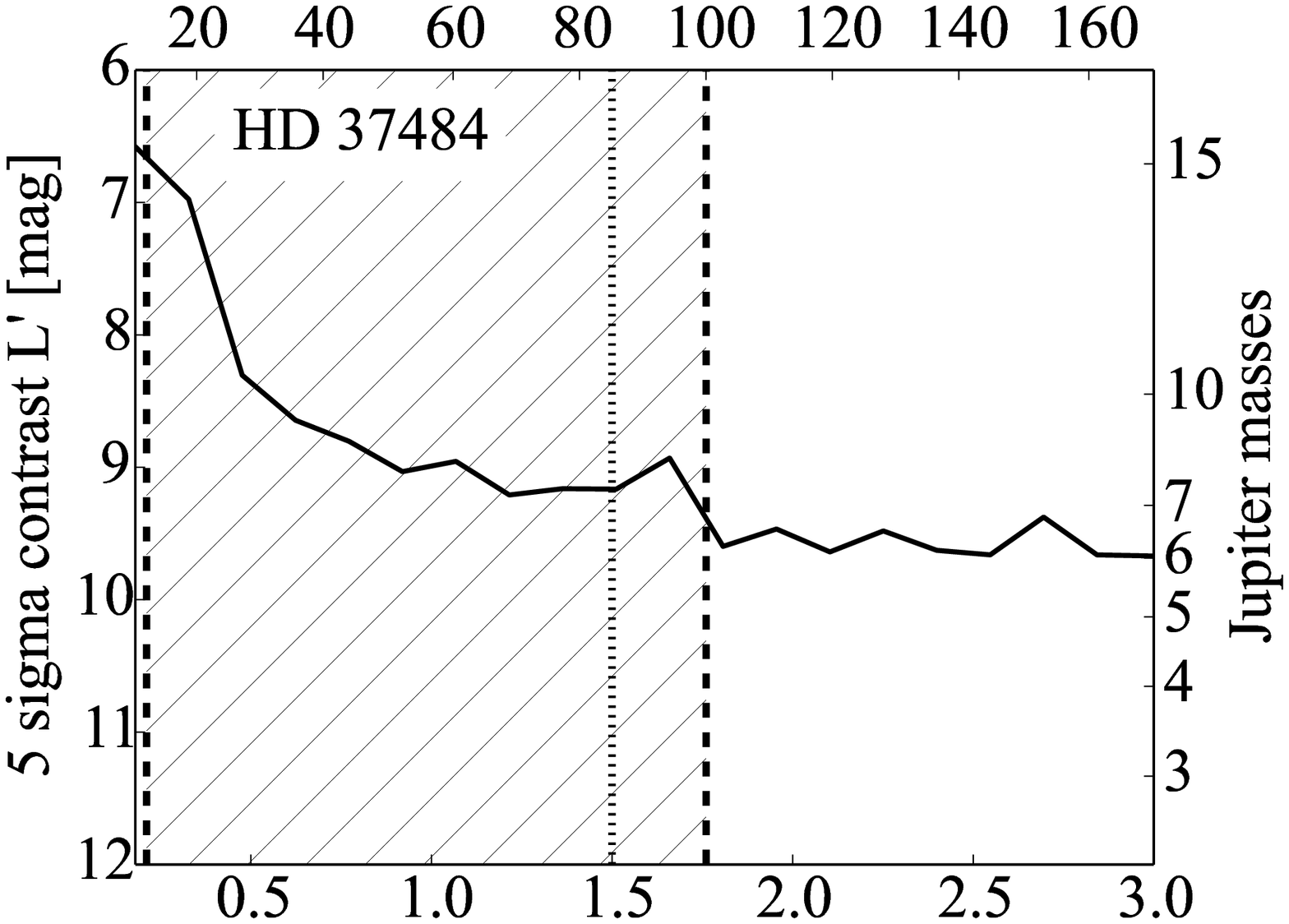}{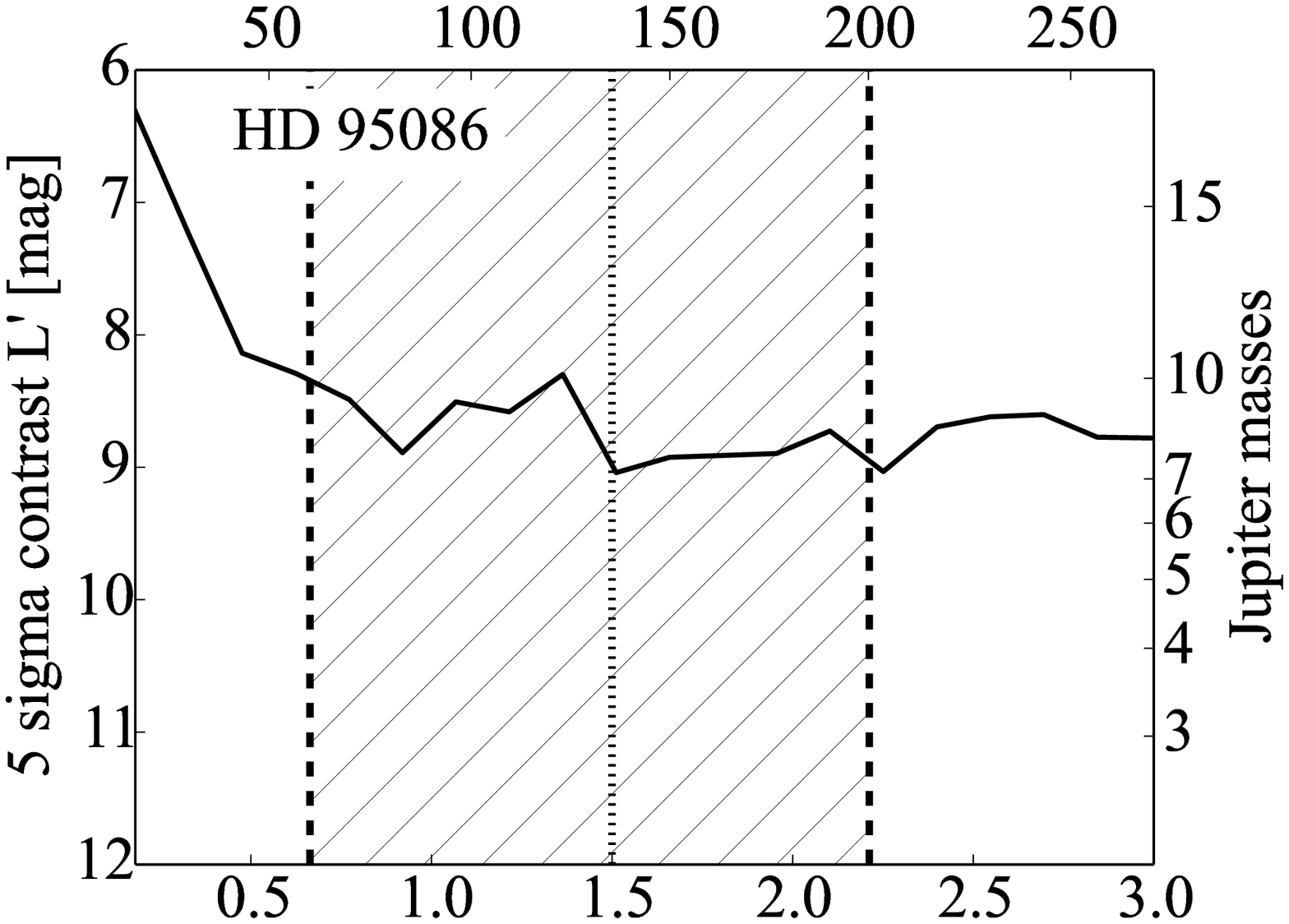}\\
\vspace{10pt}
\plottwo{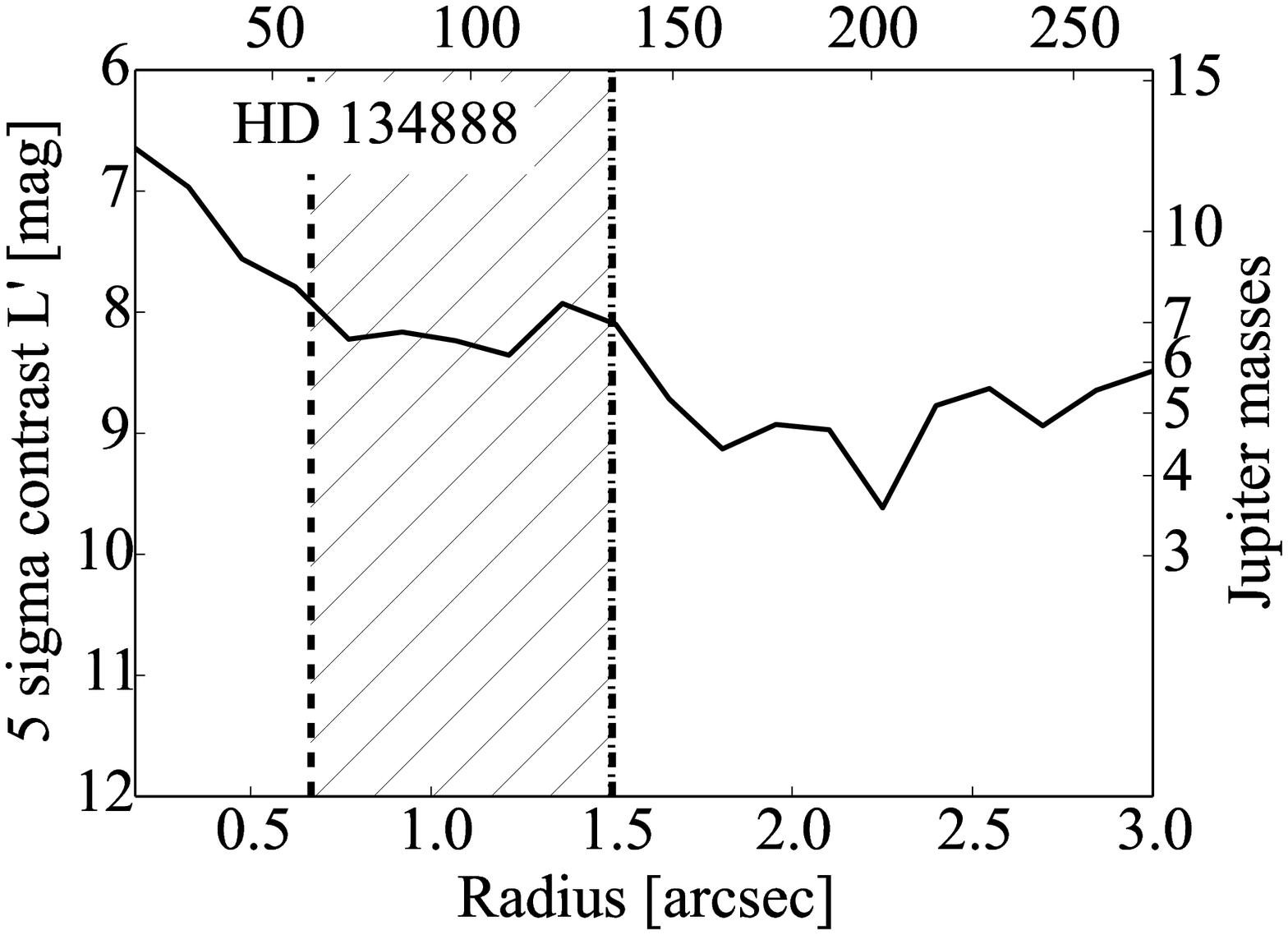}{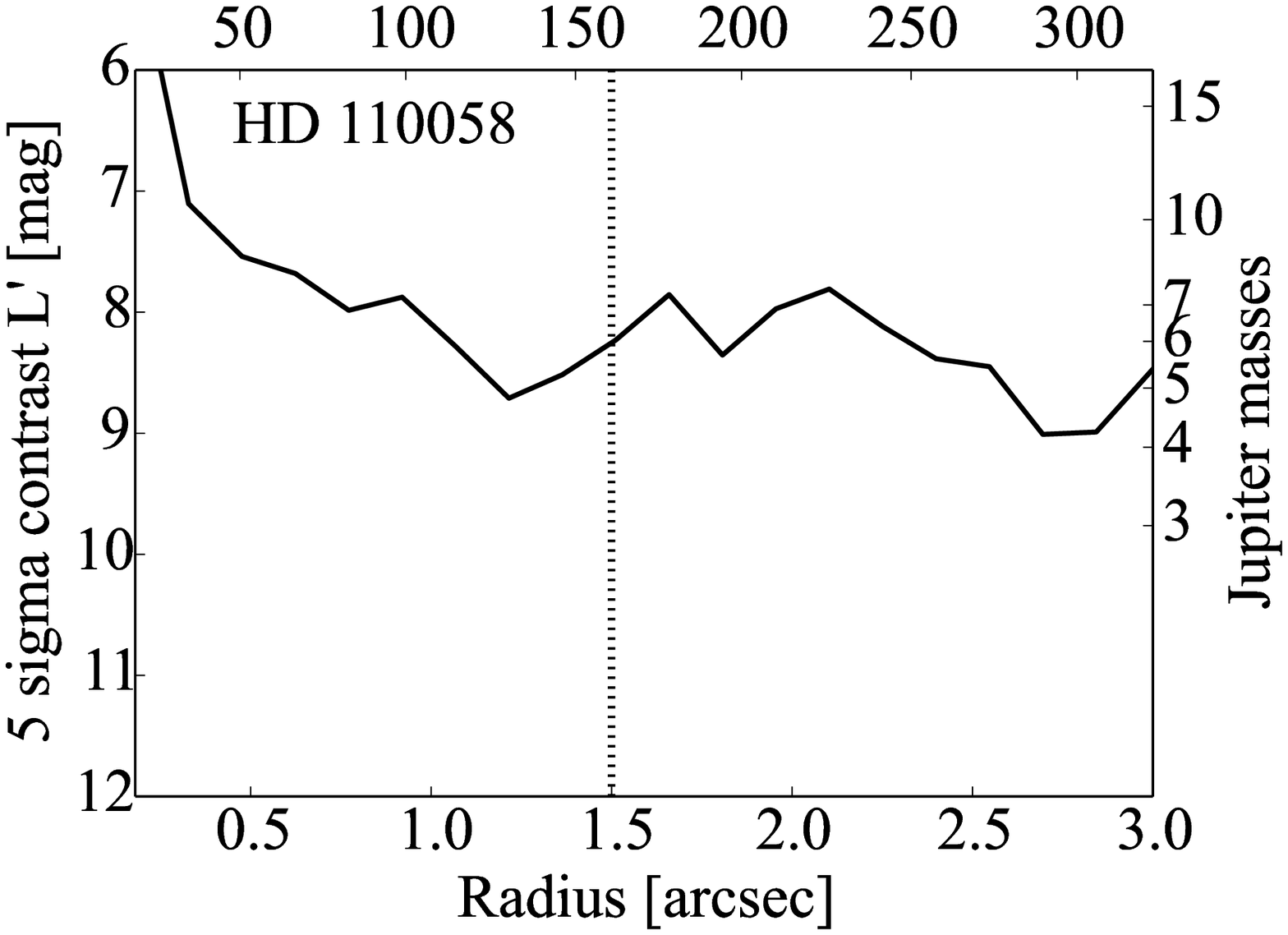}\\
\caption{Sensitivity curves for the targets. Each target has 
the 5$\sigma$ contrast curve in mag and Jupiter masses \citep[based on 
the COND evolutionary tracks][]{Baraffe03} versus separation in 
arcseconds and AU. The vertical dotted line indicates the edge of the full FOV coverage ($1\farcs5$). 
The hashed region marks the mean value of the inner and outer edge of the cold disk based on our disk modeling. 
HD 134888 and HD 110058, have only one APP hemisphere coverage. HD 134888 has 270$^{\circ}$
coverage due to large sky rotation. The contrast 
axis (left) and radius in arcsec axis (bottom) is fixed relative to each figure for comparison.}
\label{fig:cc}
\end{figure*}

The disk sizes ($R_{in}$ and $R_{out}$) estimated from our simple SED
modeling are only meant to provide a rough comparison between the
expected disk location and the point-source detection contrast curves
presented in Section \ref{sec:discussion}. Some disks we modeled
appear to have a very wide dust spatial distribution, and this is probably
because we fit the SED with only one component. It is possible that
such a cold disk SED (with a wide dust distribution) is composed of
two different populations (warm/cold belts $+$ disk halo) like the HD
95086 system (Su et al.\ in press). Furthermore, the SED from a typical debris disk,
where the sizes of dust particles are in a steep power-law form, is
dominated by small grains (less than a few times the blowout size) 
that are sensitive to non-gravitational forces. The fine
debris distribution is expected to be broader than that of their parent
bodies.

Three disks (HD 17848, HD 134888, and HD 28355) show signs of 
warm excesses in their disk SEDS (see \autoref{fig:SEDs}). In two cases (HD 17848 and HD 28355), 
the amount of warm excess is small and dominated by the errors from the
photospheric subtraction. The temperatures of the tentative excesses 
for these two systems were reported in the literature. We report the blackbody 
fit for the warm component in the remaining system, HD 134888. As the focus of this 
survey was to search for companions sculpting the cold disk components, we only 
derived detailed parameters for the cold components.

We summarize the derived disk properties: cold disk inner and 
outer radii, and dust mass (grains up to 1 mm) in \autoref{table:properties}. 
In addition to the cold disk properties, \autoref{table:properties} also lists
the distance and $L'$ mag for our sources ($L'$ mag converted from 
the 2MASS survey \citet{Cutri03} following \citet{Cox00}). The disk properties for 
HD 110058 are not calculated due to the lack of sky coverage. The distances 
are extracted from the parallax \citep{vanLeeuwen07}.

\begin{figure*}
\plotone{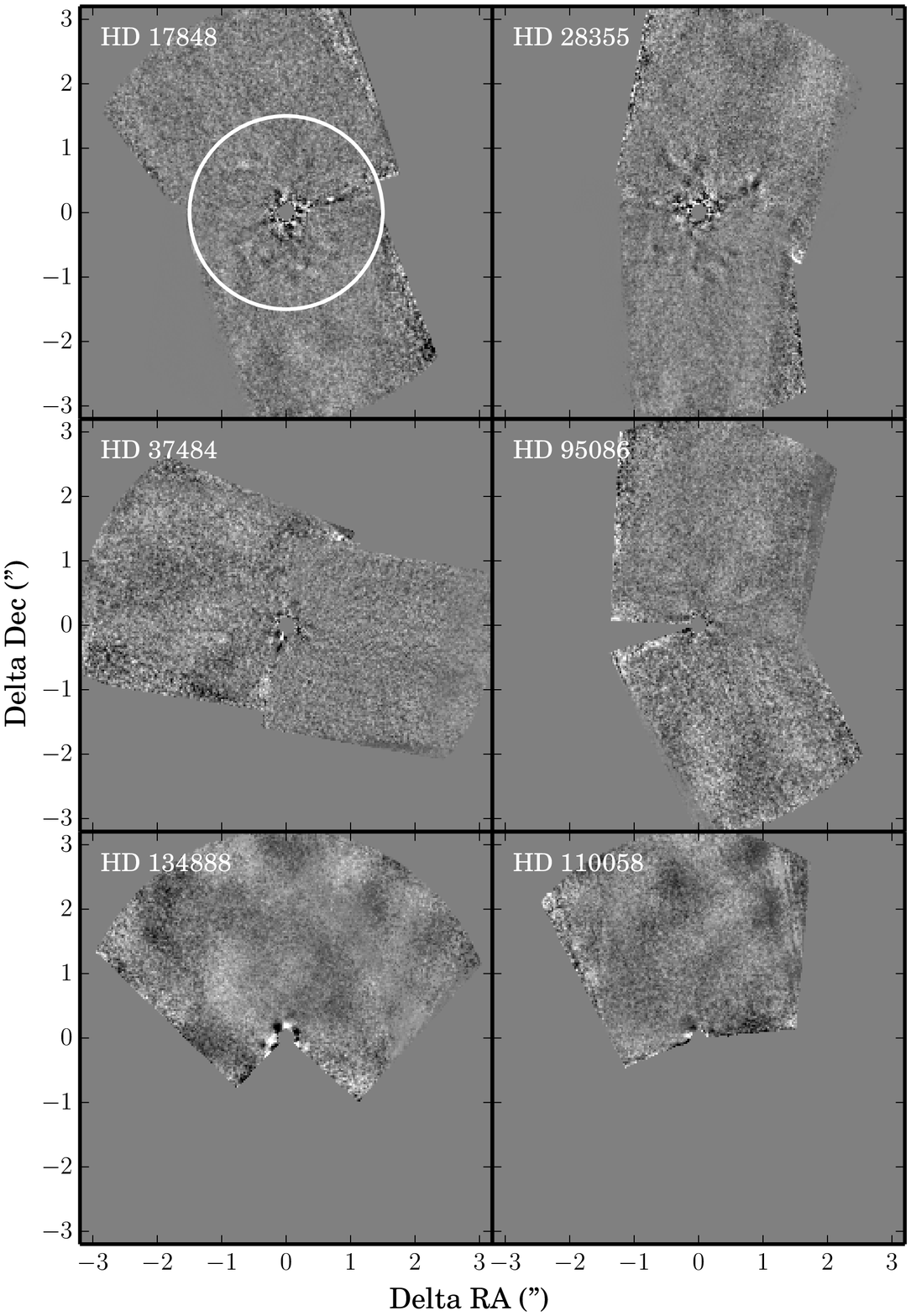}
\caption{Final PCA reduced images of all six of our targets generated with optimized PCA. 
The full sensitivity FOV is indicated by the white circle in HD 17848 at $1\farcs5$.
Beyond $1\farcs5$ we do not have full coverage because we only observe on the 
upper quartile portion of the CCD. The color range in each of the images is limited to $\pm5 \sigma$. 
The point source around HD 17848 is an artifact.}
\label{fig:HDD_images}
\end{figure*}

\autoref{fig:cc} shows the resulting 5$\sigma$ 
contrast curves for each of the targets. The dotted line is the edge of the full FOV coverage, 
beyond this line we have reduced coverage (see \autoref{fig:HDD_images}). The hashed region 
is the mean value of the inner and outer edge of the cold debris belt, based on our SED modeling. 
For each target, we adopt the age of the system (discussion below) and convert the contrast limit to 
planet masses with the COND evolutionary models \citep{Baraffe03}.
On average we are sensitive to planetary mass companions ($\lesssim$13 $M_{Jup}$) 
outward of $0\farcs5$ from the star, with the exception of 
HD 17848 and HD 28355 which are older than the other targets. 
HD 134888 and HD 110058 have only one APP hemisphere coverage. HD 134888 
achieved 75\% coverage (P.A.=135\arcdeg\ to -135\arcdeg, clock-wise passing through North) 
around the star due to large sky rotation. HD 110058 did 
not have large sky rotation (P.A.=100\arcdeg\ to -90\arcdeg, clock-wise passing through North), 
thus while the contrast curve is shown, we do not 
consider this target fully observed.
Below we discuss each of the targets individually. 

\subsection{HD17848}
HD 17848 ($\nu$ Hor) is an A2V \citep{Houk75} field star at distance
$d$ = 50.5\,$\pm$\,0.5 pc \citep[adopting parallax $\varpi$ =
19.82\,$\pm$\,0.18 mas from ][]{vanLeeuwen07}. We estimate the
effective temperature to be T$_{\rm eff}$ = 8470\,$\pm$\,130 K based on
multiple photometric T$_{\rm eff}$ estimates: using UBVK$_s$ photometry
with the color/T$_{\rm eff}$ table of \citet{Pecaut13}, and the tight
cluster of reported T$_{\rm eff}$ estimates in the literature
\citep[][]{AllendePrieto99,Paunzen06,Rhee07,Patel14}. Our adopted
T$_{\rm eff}$ is systematically cooler than that reported by
\citet{Chen14} (9000\,K), which is an outlier compared to the other
estimates. Adopting the V magnitude from \citet{Mermilliod94} (V =
5.254\,$\pm$\,0.005), \citet{vanLeeuwen07} parallax, and adopted
T$_{\rm eff}$-appropriate bolometric correction BC$_V$ =
-0.040\,$\pm$\,0.034 \citep{Pecaut13}, we estimate $\nu$ Hor's
luminosity to be log(L/L$_{\odot}$) = 1.222\,$\pm$\,0.016 dex. Based
on our HR diagram position for $\nu$ Hor, we use the evolutionary
tracks of \citet{Bertelli09} to infer an isochronal age of $\sim$540
Myr and mass 1.90\,M$_{\odot}$.  Sampling a reasonable range of
protostellar chemical compositions (Y = 0.26-0.27, Z = 0.014-0.017),
we estimate isochronal age 1$\sigma$ uncertainties of $\pm$90 Myr
(statistical) and $\pm$60 Myr (systematic), and mass 1$\sigma$
uncertainties of $\pm$0.02 M$_{\odot}$ (statistical) and $\pm$0.04
M$_{\odot}$ (systematic).  Hence, we do not find support for the young
age of $\sim$100 Myr proposed by \citet{Rhee07}, and estimate $\nu$
Hor to be $\sim$5$\times$ older than previously thought.

Its IRS spectrum was first analyzed by
\citet{Ballering13}, and suggested the system is a two-component disk
with dust temperatures of 164 K and 50 K. However, \citet{Chen14}
suggested a much warmer temperature, 353$^{+4}_{-8}$ K, while the cold
temperature (57$\pm$5 K) was consistent with the early
result. The discrepancy is most likely due to how the excess flux in 
the mid-infrared range was determined (especially how the different 
modules of IRS spectra joined and pinned down to the photosphere).
Therefore, the nature and amount of warm excess is still in
debate. Using the observed dust temperature as gauges, the warm excess
characterized by \citet{Ballering13} would be asteroid-like with
orbital distances of 9--12 AU, while the value from \citet{Chen14}
would be terrestrial-like with orbital distances of 2--3 AU. The new
far-infrared photometry from \textit{Herschel} and the MIPS-SED data
confirm that the cold disk has a typical dust temperature of $\sim$56
K (see \autoref{fig:SEDs}). We excluded data points shortward of
$\sim$30 $\mu$m in our SED model, resulting in an inner cold disk radius of
96$^{+9}_{-37}$ AU and an outer cold radius of 410$^{+24}_{-96}$ AU with a
total dust mass of (1.3$\pm$0.7)$\times 10^{-3} M_{\earth}$.

No bright point sources were detected around HD 17848 (see \autoref{fig:HDD_images}). 
The overlapping region between two APP hemispheres is noisy due to the edge 
of the dark APP hole. Therefore, the bright point (at $0\farcs7$ and P.A.=-74\arcdeg) on the reduced image 
is probably a noise spike. We compared any
possible faint point sources in our data with archival data obtained
with NaCo/VLT from both 2011 (087.C-0142(A)) and 2009 (084.C-0396(A)) 
and verified that none of these
faint point sources are present in all three datasets.  

Based on the
disk structure, we would expect to find companions inside $1\farcs9$
if the inner edge of the cold disk is maintained by
unseen planet(s). Our high-contrast observations rule out the presence
of a low-mass star down to the brown dwarf regime in the range of
0\farcs5 --3\arcsec ($\sim$25--150 AU). Bear in mind, as shown in \autoref{fig:HDD_images}, 
beyond $1\farcs5$ we do not have complete sky coverage around the star. As mentioned in 
Section \ref{subsec:ancillary_data}, the HD 17848 disk may be an edge-on. This 
limits our ability to detect companions unless they are fortuitously at a 
projected separation outwards of $0\farcs5$.

\subsection{HD 28355}
HD 28355 is an A7V \citep{Abt95} member of the 650 Myr-old Hyades cluster
\citep{Su06,DeGennaro09} at a distance of 48.8\,$\pm$\,0.6 pc away \citep{vanLeeuwen07}. 
Its IRS spectrum along with MIPS 70 $\mu$m
photometry has been published by several papers with different derived
dust temperatures. \citet{Morales11} suggest the disk SED is best
described by two temperatures of 128 K and 60 K, which are consistent
with the values (120 K $+$ 54 K) published by \citet{Ballering13}
within errors. Analysis performed by \citet{Chen14} gives the warm
temperature of 176$^{+7}_{-8}$ K and the cold temperature of
69$^{+5}_{-6}$ K. As discussed in Section \ref{sec:disk_properties}, 
HD 28355 is near a bright infrared 
source, which is extended at 160 $\mu$m, as a result its far-infrared
fluxes are contaminated. 
The overall disk SED can be fit with two
temperatures: $\sim$120 K and $\sim$55 K (see \autoref{fig:SEDs}). However, the
contamination-free part of the SED (data shortward of $\sim$80 $\mu$m)
can also be described by one single temperature of $\sim$80 K (within
a few $\sigma$). Therefore, we excluded the \textit{Herschel} 160
$\mu$m data and the MIPS-SED points longward of 85 $\mu$m in our
one-component SED model. The inner boundary is estimated to be
46$\pm$12 AU, and the outer is 130$\pm$30 AU with a total dust mass of
(1.8$\pm$0.7)$\times10^{-3} M_{\earth}$.

No bright point sources were detected around HD 28355 (see \autoref{fig:HDD_images}). The 
structure inwards of $1\farcs5$ is due to residuals from the APP PSF. We were not sensitive to planetary mass
companions due to the older age of the star. We were able to rule out substellar 
companions from 0\farcs5 --3\arcsec ($\sim$24--147 AU). Our single component SED model 
places the inner boundary of the disk at $0\farcs9$. Thus, 
if the cold disk is shaped by a companion, it must be less than $\sim$50 $M_{{Jup}}$.

\subsection{HD 37484}

HD 37484 is a F3V \citep{Houk82} star at a distance of 
$d$ =  56.8\,$\pm$\,2.0 pc \citep[adopting parallax $\varpi$ =
17.61\,$\pm$\,0.62 mas from ][]{vanLeeuwen07}. Its $B-V$ vs. $M_{V}$ position 
places it on the zero age main sequence (ZAMS), thus it must 
be $>$27 Myr in order to allow enough time for a $\sim$1.36 $M_{\odot}$ star to
reach the main sequence. Its position is 
commensurate with that of the Pleiades and IC2391 clusters, making it $<100$ Myr.
It has a brighter and hotter ($\sim$A5V) common
proper motion companion 0.34 deg away (HR 1915). Both stars are
considered Columba members by \citet{Malo13}, who list 10-40 Myr
for the group.  The companion appears to be on or below the ZAMS, consistent
with an age of $>$19 Myr. HD 37484's HR diagram position is 
consistent with the previous assigned age of $\sim$30 Myr. Given its 
Columba membership and consistency with the age of other Columba members, 
we adopt the age of $30\,\pm\,10$ Myr.

The disk SED is composed of data from
\textit{Spitzer} MIPS 24 and 70 $\mu m$ photometry as well as IRS
10--35 $\mu m$ spectroscopy. The disk SED is best fit with a blackbody temperature
of $\sim$88 K.  The best fitting disk model has an inner radius of
$12^{+20}_{-4}$ AU, outer radius of $100^{+100}_{-20}$ AU and dust
mass of (2$\pm$1)$\times10^{-3}$ M$_{\earth}$.  This source does not
have longer wavelength {\it Herschel} data, which makes the outer disk
radius and dust mass fairly unconstrained.

We initially detected a faint point source around HD 37484 at
$0\farcs97$ (55 AU from the star), 9.0 magnitudes fainter than the host star at a
P.A. of 103$^{\circ}$ in our APP data (see \autoref{fig:HDD_images}).  However, this point source was
not detected in archival data with equivalent sensitivity taken in
2011 (088.C-0085(A)), allowing us to conclude that it is unlikely to
be a real source. This false detection demonstrates the importance of
archival data, which can be used to quickly confirm or deny the
physical nature of a point source.

\subsection{HD 95086}
HD 95086 is a 17 Myr old A8 LCC member star at a distance of 90.4\,$\pm$\,3.3 pc \citep{Meshkat13}. 
The system was also resolved by \textit{Herschel} at 70 and 100 $\mu$m 
\citep{Moor13} with an estimated inclination of 25\arcdeg\ from face-on. 
Re-analysis of the \textit{Herschel} resolved images combined with 
detailed SED modeling reveal that the extended part of the images 
arises from a disk halo (Su et al.\ in press), similar to the disk halo found 
around HR 8799 (\citealt{Su09, Matthews14}). In the 
three-component disk model presented by Su et al.\ (in press), the inner 
belt is located from $\sim$7 AU to $\sim$10 AU, and the cold planetesimal 
disk likely ranges from 63$\pm$6 AU to 189$\pm$13 AU, and are 
surrounded by an extended halo up to $\sim$800 AU. 

A 5 $M_{{Jup}}$ planet was discovered around HD 95086 by
\citet{Rameau13b}. Our $H$-band NICI/Gemini non-detection provided a strict color lower limit 
 for the planet of $H-L'>3.1\,\pm\,0.5$ mag \citep{Meshkat13}. 
 The subsequent $H$-band Gemini Planet Imager (GPI) detection of HD 95086 b by \citep{Galicher14}
provides a red planet color of $H-L'=3.6\,\pm\,1.0$. We also observed the system with the APP, but did not
detect it due to decreased AO quality at high airmass ($>$\,1.4) during the observation (APP hemisphere 2, 
see \autoref{fig:HDD_images}). Due to lack of sky rotation, we are missing about 20$^{\circ}$ 
of coverage. \autoref{fig:cc} shows that we achieve a sensitivity of $\sim$10 $M_{{Jup}}$ at the 
separation of the planet, however the detected planet is $5\,\pm\,2$ $M_{{Jup}}$  \citep{Rameau13c}.

\subsection{HD 134888}

HD 134888 is an F4V star, located 90$^{+9}_{-8}$ pc away \citep[adopting parallax $\varpi$ =
11.12\,$\pm$\,1.05 mas from ][]{vanLeeuwen07}. 
We adopt an age of 16 Myr based on its membership in the Lower Centaurus-Crux 
(LCC) association, commensurate with isochronal age estimates for the star 
(16-25 Myr; \citealt{Pecaut12}). \citet{Chen14} conclude it has a two-component disk
with the warm temperature of 387$^{+6}_{-7}$ K and the cold
temperature of 72$\,\pm\,$5 K. The far-IR excesses along with the IRS
excess longward of $\sim$30 $\mu$m are consistent with a blackbody
temperature of 75 K (see \autoref{fig:SEDs}). However, we found the
warm excess is better fit with a dust temperature of $\sim$175 K
(i.e., asteroid-like). The derived temperature in the warm component is highly dependent on the
amount and shape of the excess emission in the 10--20 $\mu$m
region. Future mid-IR observations are needed to better characterize the
warm component. We excluded data points shortward of $\sim$30 $\mu$m
in our SED fitting of the cold component. The best-fit one cold-component disk ranges from
$60\pm11$ AU to $135\pm29$ AU with a dust mass of
$3.25\,\pm\,1.4\times10^{-2}$ M$_{\earth}$.

Based on our inferred disk structure, the ideal place for detecting the 
potential low-mass companion that sculpts the disk is interior of 
60 AU (0\farcs66) or exterior of 135 AU (1\farcs5).  
A possible interesting point source appeared at
$0\farcs2$ from the star, but after comparing with archival
data (087.C-0142(A)), the point source appears to be an artifact of the
data reduction. No believable point sources are detected in our reduced images around the 
star at any projected separation, however we cannot
conclusively say there are no companions around HD 134888. 
From a P.A. of 135$^{\circ}$ to -135$^{\circ}$ (clock-wise, passing through
North, see \autoref{fig:HDD_images}), we do not detect any companions with an average mass limit of
5-7 $M_{{Jup}}$.

\subsection{HD 110058} 

HD 110058 is a A0V star \citep{Houk78} and member of the LCC 
subgroup of Sco-Cen \citep{deZeeuw99,Rizzuto11}.
The revised Hipparcos parallax from \citet{vanLeeuwen07} ($\varpi$ =
9.31\,$\pm$\,0.78 mas) translates to a distance of 107$^{+10}_{-8}$
pc. The star is situated near $\ell$, $b$ = 301$^{\circ}$,
+13$^{\circ}$.6, in the northern part of LCC, which appears to be the
oldest part of the subgroup \citep{Preibisch08,Pecaut12}, so we follow
\citet{Chen14} and adopt an age of 17 Myr.

We did not achieve full $360^{\circ}$ APP coverage around this star . Only one APP hemisphere was 
observed and it had poor sky rotation. Without full sky coverage, the contrast curve for HD 110058 
(\autoref{fig:cc}) is only valid from a P.A. of $\sim100^{\circ}$ to -90$^{\circ}$ (clock-wise, passing through
North, see \autoref{fig:HDD_images}). We are sensitive to 5--8 $M_{{Jup}}$ from 0\farcs5 --3\arcsec. We detect no point sources in 
this limited region around the star.

\section{Conclusions}
\label{sec:conclusions}
We present the first results from our survey searching for planets around 
stars with bright debris disks with gaps. In this paper, we present 
the data from six targets obtained with the APP coronagraph on NaCo/VLT. 
One of our targets, HD 95086, was found to harbor a planet 
on the inner edge of the outer debris belt \citep{Rameau13b, Moor13, Galicher14}. 
While our data were not sensitive enough to detect this planet \citep{Meshkat13}, its discovery 
demonstrates the strength of two-component disk targets.

For each of the targets, we derive disk properties based on our  
SED models. These properties (including disk mass 
and debris radial distribution) allow us to infer the likely location of gaps in the disk, which 
may be carved out by planets. We detect no companions in our APP data. 
A few plausible point sources were detected but 
ruled out after comparison with archival data. We were sensitive to planetary-mass 
companions for four of the six targets, using COND atmosphere models. 
If cool planets have substantial methane absorption and little cloud opacity, as 
is predicted by the COND evolutionary models \citep{Baraffe03}, then L'-band will 
be less sensitive to planets. However, the majority of directly imaged planets do 
not behave like field brown dwarfs of similar effective temperature. Most cool 
planets do not show evidence of methane absorption (eg: HR8799bcd and 2M1207b \citealt{Skemer14}) 
and are redder than predicted ($H-L'$ = 2 to 3 mag: \citealt{Galicher14}) Thus, our use 
of L'-band is comparable to the best H-band surveys at separations where we are contrast limited.
The benefit of the APP for companion discovery 
over direct imaging is inconclusive, based on our sample of six observations. 
See Meshkat et al. (in prep.) for further discussion.

The sample size of six targets is too small
to draw conclusions about the origins of the gaps in Holey Debris Disks. 
Our complete Holey Debris Disks sample (Bailey et al. in prep.) will allow 
stronger statements of whether the gaps in these disks are formed by one massive companion,  multiple 
low-mass companions, or other mechanisms.
We discuss each target individually and analyze the sensitivity 
of companions achieved relative to the boundaries of the debris disks,
based on our disk models. 

In order to detect lower mass planets at the inner edges of the cold outer debris belts, greater 
sensitivity needs to be achieved. In this paper, we have only 
modelled the outer single cold component, however many of our targets may be two-component disks. 
The projected separation of the warm inner disk components are much less than the 
inner working angle limit of current high-contrast imaging data, and so planets sculpting 
the warm disks would remain undetected. Discoveries like that of the low-mass planet
HD 95086 b strengthen the notion that gaps need not be carved by a single very massive 
companions \citep{DodsonRobinson11}, and thus future surveys will require increased sensitivity 
in addition to smaller inner working angle.

\acknowledgements{We thank the anonymous referee for their suggestions
which improved this paper. We thank S. Quanz for assistance in acquiring these 
data. TM and MAK acknowledge funding
under the Marie Curie International Reintegration Grant 277116
submitted under the Call FP7-PEOPLE-2010-RG. VB acknowledges support 
from the NSF Graduate Research Fellowship Program (DGE-1143953).
KYLS is partially supported by NASA grant \#NNX11AF73G. 
EEM acknowledges support from NSF award AST-1313029.
This paper made use of the SIMBAD and VIZIER databases.}

\bibliographystyle{apj}

\end{document}